\documentclass[%
 aip,%
 cp,%
 amsmath,amssymb,%
 reprint,%
 groupedaddress,
]{revtex4-1}

\usepackage[dvipsnames]{xcolor}
\usepackage[colorlinks=true,bookmarks=false,linkcolor=NavyBlue,urlcolor=NavyBlue,citecolor=NavyBlue,breaklinks]{hyperref}
\usepackage{amsmath}
\usepackage{amsfonts}
\usepackage{graphicx}
\usepackage{color}
\usepackage{braket}
\usepackage{graphicx}
\usepackage{dcolumn}
\usepackage{bm}

\begin{document}
\preprint{AIP/123-QED}
\title{Effect of surface hybridization on RKKY coupling in ferromagnet/topological insulator/ferromagnet trilayer system}

\author{Cong Son Ho}
 \email{elehcs@nus.edu.sg}
 \affiliation{
Department of Electrical and Computer Engineering, National University of Singapore,4 Engineering Drive 3, Singapore 117576, Singapore.
}
\author{Mansoor B. A. Jalil}%
 \email{elembaj@nus.edu.sg}
\affiliation{
Department of Electrical and Computer Engineering, National University of Singapore,4 Engineering Drive 3, Singapore 117576, Singapore.
}

\date{\today}

\begin{abstract}
We theoretically investigate the RKKY exchange coupling between two ferromagnets (FM) separated by a thin topological insulator film (TI). We find an unusual dependence of the RKKY exchange coupling $\Phi_\mathrm{ex}$ on the TI thickness ($t_{TI}$). First, when $t_{TI}$ decreases, the coupling amplitude increases at first and reaches its maximum value at some critical thickness, below which the amplitude turns to diminish. This trend is attributed to the hybridization between surfaces of the TI film, which opens a gap below critical thickness and thus turns the surfaces into insulating state from semi-metal state. In insulating phase, diamagnetism induced by the gap-opening compensates paramagnetism of Dirac state, resulting in a diminishing magnetic susceptibility and RKKY coupling. For typical parameters, the critical thickness in $\mathrm{Bi_2Se_3}$ thin film is estimated to be in the range of 3-5 nm.
\end{abstract}

\pacs{75.70.Cn, 03.65.Vf, 73.40.-c}
\maketitle

\section{Introduction}

In the past decade, topological insulators (TI) have emerged as one of the most attractive topics, in both theory and experiment \cite{Kane:prl05,Bem:sci06,Xia:nat09,Mell:nat14,Wang:prl15,Fan:nat14}. Topological insulators possess surface states with strong spin-momentum locking \cite{Kane:prl05,Bem:sci06,Xia:nat09}, which translates to a large spin-dependent effects such as spin-orbit torques \cite{Mell:nat14,Wang:prl15,Fan:nat14} and quantum anomalous Hall effects \cite{Chang:sci13,Qi:prl16}. Moreover, the spin-locked topological surface states of TI are robust under effects of the time-reversal symmetric impurity scattering. These factors enable TI to be one of the most promising candidates for spintronic devices \cite{ga:prl10}.

At the same time, ferromagnetic hetero-structures (FMs) are also integral spintronic elements that have been used as a platform for practical spintronic devices, such as spin Hall \cite{Jung:nat12,Wunder:nat09,Wunder:sci10,Seki:nat08} and  spin-orbit-based memories  \cite{Jalil:srep14,Kent:nat15}, topological Hall-based sensors  \cite{Ni:ieee16}. When two ferromagnets form a spin-valve structure where they are separated by a metal spacer, there is a interlayer exchange coupling between the FMs mediated by the itinerant electron in the metal spacer, which is known as RKKY coupling \cite{bruno1995theory,bruno1993interlayer}. The RKKY coupling oscillates with spacer thickness between ferromagnetic and antiferromagnetic values, and typically its amplitude decreases with increasing thickness. Recently, the RKKY coupling has been studied in TI systems \cite{liu2009magnetic,li2015magnetic}, where the magnetic moments can be in the same surface \cite{liu2009magnetic} or belong to different surfaces \cite{li2015magnetic} of the TI film. In the latter case \cite{li2015magnetic}, by assuming unchanged surface states as the thickness changes, the interlayer was shown to behave in the same way as in FM/metal/FM system, i.e., oscillatory damping with increasing TI thickness. However, in TI films which are very thin, coupling between surfaces becomes significantly dependent on the film thickness \cite{Linder:prb09,Lu:prb10} and it can critically modify the surface states and their magnetic property \cite{Zyu:prb11}.

In this work, we study the interlayer coupling between two FMs separated by a thin topological insulator film \cite{luo2013massive, wang2015electrically, li2015magnetic}, where thickness-dependence of the surface states is considered. We find that, besides the oscillatory behavior, the coupling amplitude is not monotonically dependent on the thickness, but there is a critical value of thickness at which the coupling amplitude reaches its peak. Below the critical value, the coupling does not increase but rather decreases. These trends are attributed to the phase transition at the critical thickness \cite{Zyu:prb11} and strong surface hybridization in thin TI films.

\section{Theory}

\begin{figure}
	\includegraphics[width=0.4\textwidth]{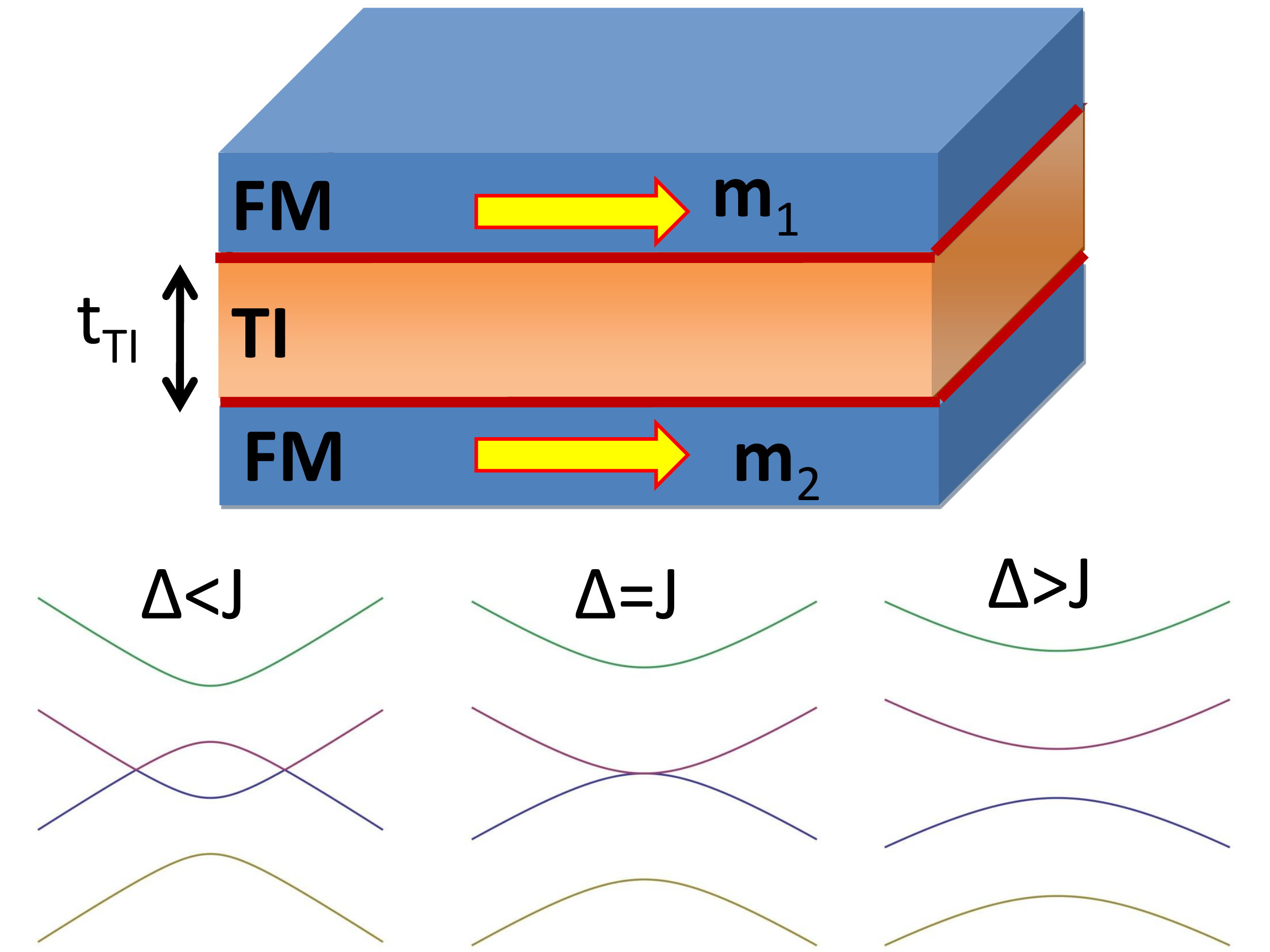}
 \caption{(Color online) Schematic diagram of ferromagnet/topological insulator/ferromagnet (FM/TI/FM) trilayer. $t_{TI}$ is the thickness of the TI thin film. Bottom panel shows the energy bands for various value of the inter-surface coupling $\left(\mathrm{\Delta }\right)$, with $J$ being the value of the sd coupling between electron spin and FM. \label{Fig1}}
\end{figure}

\noindent We consider a trilayer system comprising of a thin topological insulator (TI) film sandwiched by two ferromagnetic films (see Fig. \ref{Fig1}). In this structure, both top and bottom Diract surface of the TI film are active, and they are independently coupled to top and bottom FMs, respectively, via $sd$ coupling. Note that in a super-thin film, the top FM can also couple to the bottom TI surface and vice versa, however, for the simplicity such couplings are neglected. The system is described by model Hamiltonian
\begin{equation} \label{GrindEQ__1_} 
\mathcal{H}=\left[ \begin{array}{cc}
h_{TI}+J{\bf{m}}_{\bf{1}}\cdot\hat{\sigma } & \mathrm{\Delta }I_2 \\ 
\mathrm{\Delta }I_2 & -h_{TI}+J{\bf{m}}_2\cdot\hat{\sigma } \end{array}
\right],                                
\end{equation} 
where $h_{TI}={\hbar v}_F\left(~\hat{z}\times \hat{\sigma }\right)\cdot\bf{k}$ describes the effective Hamiltonian of the (top) topological surface state of the TI film, with $v_F$ being the Fermi velocity, $\hat{\sigma}$ is the vector of the Pauli matrices, and $\hat{z}$ being the normal unit vector. Whereas, the bottom surface state, which has opposite helicity, is then described by $h_{TI}(-{\bf{k}})=-h_{TI}(\bf{k})$.  The magnetizations are assumed to align along in-plane directions to assure that the gap opening is solely induced by the surface hybridization, see later. The off-diagonal elements describe the hybridization between top and bottom TI surfaces (inter-surface coupling), which is quantified by the tunneling element  $\mathrm{\Delta }$. In free-standing TI films, the hybridization opens a gap of $2\mathrm{\Delta }$ in the TI surface states. In previous work \cite{Zyu:prb11}, it has been shown that the gap can be close or open by applying an appropriate in-plane magnetic field. In our case, the proximity coupling ($J\mathbf{m}$) can take the role of driving field to adjust the gap, see Fig. \ref{Fig1}.

 In thick TI films, the hybridization coupling is vanishingly small, meanwhile in the limit of thin TI films, it can be related to the TI thickness  $t_{TI}$ as  \cite{Linder:prb09,Lu:prb10}
\begin{equation} \label{GrindEQ__2_} 
\mathrm{\Delta }\approx \frac{{\pi }^2B_1}{t^2_{TI}},         
\end{equation} 
with $B_1$ is a material-dependent parameter. With typical value of $B_1=0.1\ \mathrm{eV}\ \mathrm{nm}^2$ in $\mathrm{Bi_2Se_3}$ \cite{Zhang:nat10,Liu:prb10}, the tunneling element is $0.05\ \mathrm{eV}$  for 5 nm film, and can be up to 0.25 eV for 2nm thin film \cite{Peng:nat10}. 

Eigenenergies of \eqref{GrindEQ__1_} are given by
\begin{equation} \label{GrindEQ__3_} 
E_{s\tau}=s \sqrt{U+\tau V} ,     
\end{equation} 
with
\begin{eqnarray}
U=J^2+\Delta^2+v_F^2\hbar^2k^2+v_F\hbar({\bf{m}}_1-{\bf{m}}_2)\cdot({\bf{k}}\times\hat{z}),\nonumber\\
V=J\sqrt{v_F^2\hbar^2[({\bf{m}}_1+{\bf{m}}_2)\cdot({\bf{k}}\times\hat{z})]^2+\Delta^2({\bf{m}}_1+{\bf{m}}_2)^2},\nonumber
\end{eqnarray}
where $s,\tau =\pm 1$ representing spin and hyperbola indices, respectively. The band diagrams are shown in Fig. \ref{Fig1}.

To derive the interlayer exchange coupling between the ferromagnets, we apply the RKKY formula given by \cite{bruno1995theory,bruno1993interlayer}
\begin{equation} \label{GrindEQ__4_} 
\Phi_\mathrm{ex}=-N\int^{k_{F\bot}}_{-k_{F\bot}}{dq_ze^{iq_zt_{TI}}\chi \left(q_{\parallel }=0,q_z\right)}.     
\end{equation} 
In the above, $k_{F\bot}$ is the Fermi momentum in the direction perpendicular to the film, which has typical value of the order of $1/a_0$, with $a_0\approx$ 1 nm being the thickness of a quintuple layer. $\chi (\bf{q})$ is the $\bf{q}$-dependent magnetic susceptibility of the TI film. $N=\frac{1}{2}{\left(\frac{A}{V_0}\right)}^2\left(\frac{S^2a^2_0}{2\pi V_0}\right)$, where $A$ is the contact potential between electron spin and ferromagnets, $S$ is the spin of the FM spin, $V_0$ is the atomic volume.

The magnetic susceptibility of the TI thin film is given by the Kubo's formula \cite{white2007quantum}
\begin{equation} \label{GrindEQ__5_} 
{\chi }_\mathrm{spin}({\bf{q}})=\frac{{g^2_s\mu}^2_B}{2}\sum_{m^>,n^<}{\int{\frac{d\bf{k}}{(2\pi)^2}\frac{f_0\left(E_{n,\bf{k}}\right)-f_0\left(E_{m,\bf{k}\bf{-}\bf{q}}\right)}{E_{n,\bf{k}}-E_{m,\bf{k}\bf{-}\bf{q}}+i0^+ }}},     
\end{equation} 
in which $m^>$ ($n^<$) are for occupied (empty) bands,  $g_s=2$ is the g-factor of electron spin, $f_0\left(E_{n,k}\right)$ is the Fermi distribution function corresponding to eigen-energy branch $E_{n,k}$. In addition, the orbital angular momentum can also contribute to the total susceptibility, however, for the simplicity we will ignore this contribution. 

\section{Results}

\noindent Substituting the energy in Eq. \eqref{GrindEQ__3_} to \eqref{GrindEQ__5_} and assuming that the Fermi energy $E_F=0$, the magnetic susceptibility at temperature $T=0$ can be evaluated as
\begin{equation} \label{X1} 
\chi_\mathrm{spin}=\frac{{\mu }^2_B}{2{\pi }^2}\frac{\left(J^2+{\mathrm{\Delta }}^2+3{\left|J^2-\mathrm{\Delta }^2\right|}\right)}{\left(J+\mathrm{\Delta }+\left|J-\mathrm{\Delta }\right|\right){\left|J^2-{\mathrm{\Delta }}^{\mathrm{2}}\right|}},        
\end{equation} 
for parallel magnetizations $\bf{m}_1\|\bf{m}_2$, and
\begin{equation} \label{X2} 
\chi_\mathrm{spin}=\frac{\mu^2_B}{\pi^2}\frac{1}{\sqrt{J^2+\Delta^2}},        
\end{equation} 
for the opposite case $\bf{m}_1\|-\bf{m}_2$. 
	\begin{figure}[t]
	\includegraphics[width=0.5\textwidth]{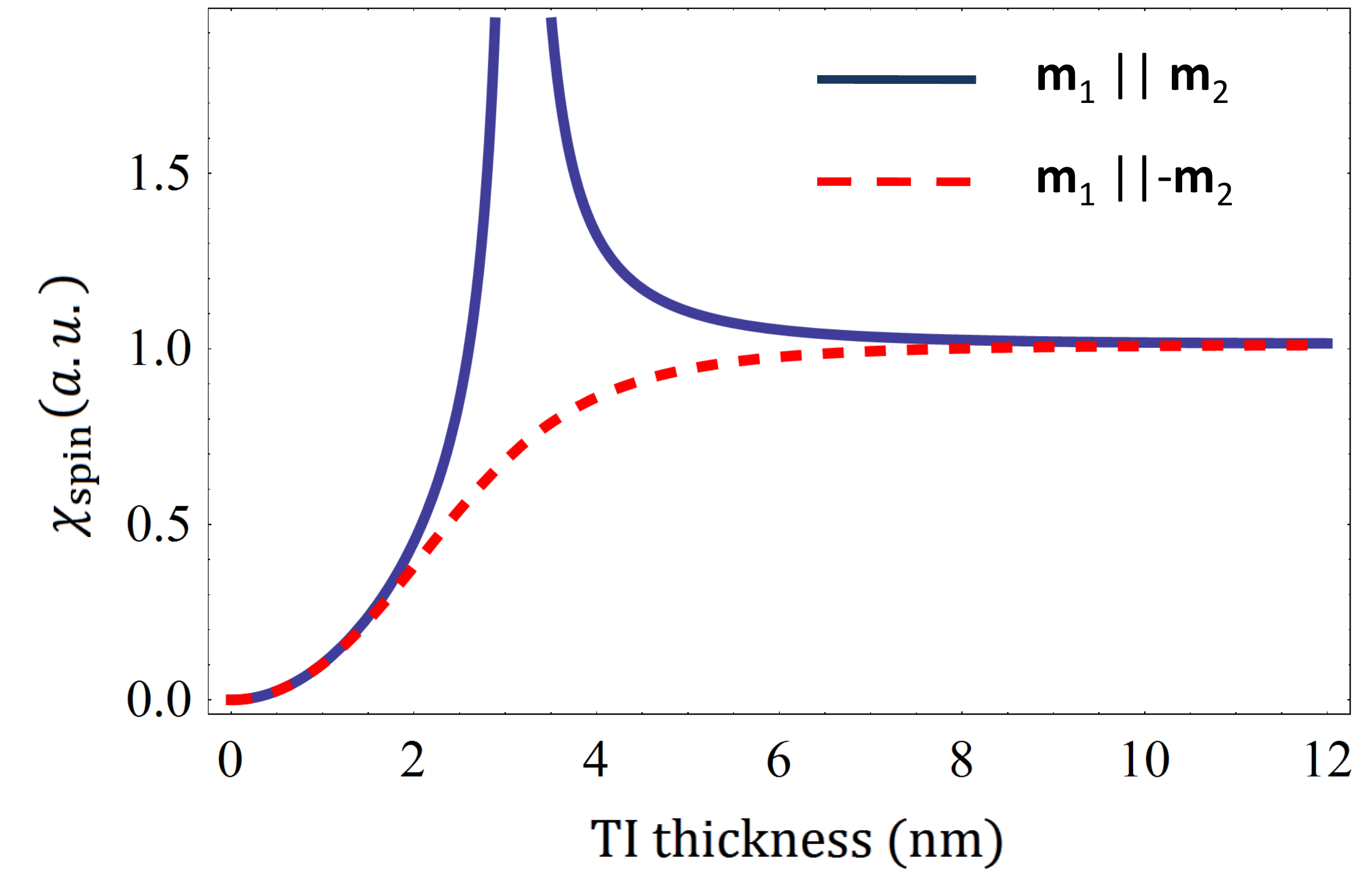}
 \caption{(Color online) The spin susceptibility of the TI film as a function of the film thickness. The singularity occurs at a critical thickness where the bands are met (see Fig. \ref{Fig1}). Other parameters: $J=0.1$ eV, $B_1= 0.1\ \mathrm{eV\ nm^2}$ \label{Fig2}}
\end{figure}

In derivation of the above equations, we have assumed strong exchange limit, i.e., $J\gg \hbar v_Fk_F$.  Otherwise, since the interlayer exchange coupling is quadratic in the proximity coupling between electron spin and FM spin (see Eq. \ref{GrindEQ__4_}), the weak FM/TI coupling limit is not interesting. From Eq. \ref{X1}, it is obvious that the spin susceptibility encounters singularity when $\Delta=J$, see Fig. \ref{Fig2}. In general, any singularity in the susceptibility relates to a second order phase transition \cite{white2007quantum}, which in this case is the transition between insulating state ($\Delta>J$) and metallic state ($\Delta<J$) at the singularity point \cite{Zyu:prb11}. The diminishing susceptibility below the critical thickness is possibly due to the emergence of diamagnetism in the insulating surface state \cite{Zyu:prb11}.  On the other hand, if the magnetizations are in the anti-parallel configuration, the band gap is always open and the surface states are always in the insulating phase, the singularity is thus avoided.

 From Eqs. \eqref{GrindEQ__4_} and \eqref{X1}, \eqref{X2} the interlayer exchange coupling is readily obtained as
\begin{equation} \label{GrindEQ__7_} 
\Phi_\mathrm{ex}={-I_0\frac{\left(J^2+{\mathrm{\Delta }}^2+3{\left|J^2-\mathrm{\Delta }^2\right|}\right)}{\left(J+\mathrm{\Delta }+\left|J-\mathrm{\Delta }\right|\right){\left|J^2-{\mathrm{\Delta }}^{\mathrm{2}}\right|}}}{\frac{2{\mathrm{sin} \left(k_{F\bot} t_{TI}\right)}}{t_{TI}}},               
\end{equation} 
for parallel magnetizations $\bf{m}_1\|\bf{m}_2$, and
\begin{equation}
\Phi_\mathrm{ex}=-I_0\frac{1}{\sqrt{J^2+\Delta^2}}{\frac{2{\mathrm{sin} \left(k_{F\bot} t_{TI}\right)}}{t_{TI}}},        
\end{equation} 
for the opposite case $\bf{m}_1\|-\bf{m}_2$, where $I_0=N \frac{\mu^2_B}{\pi^2}$.

	\begin{figure}
	\includegraphics[width=0.5\textwidth]{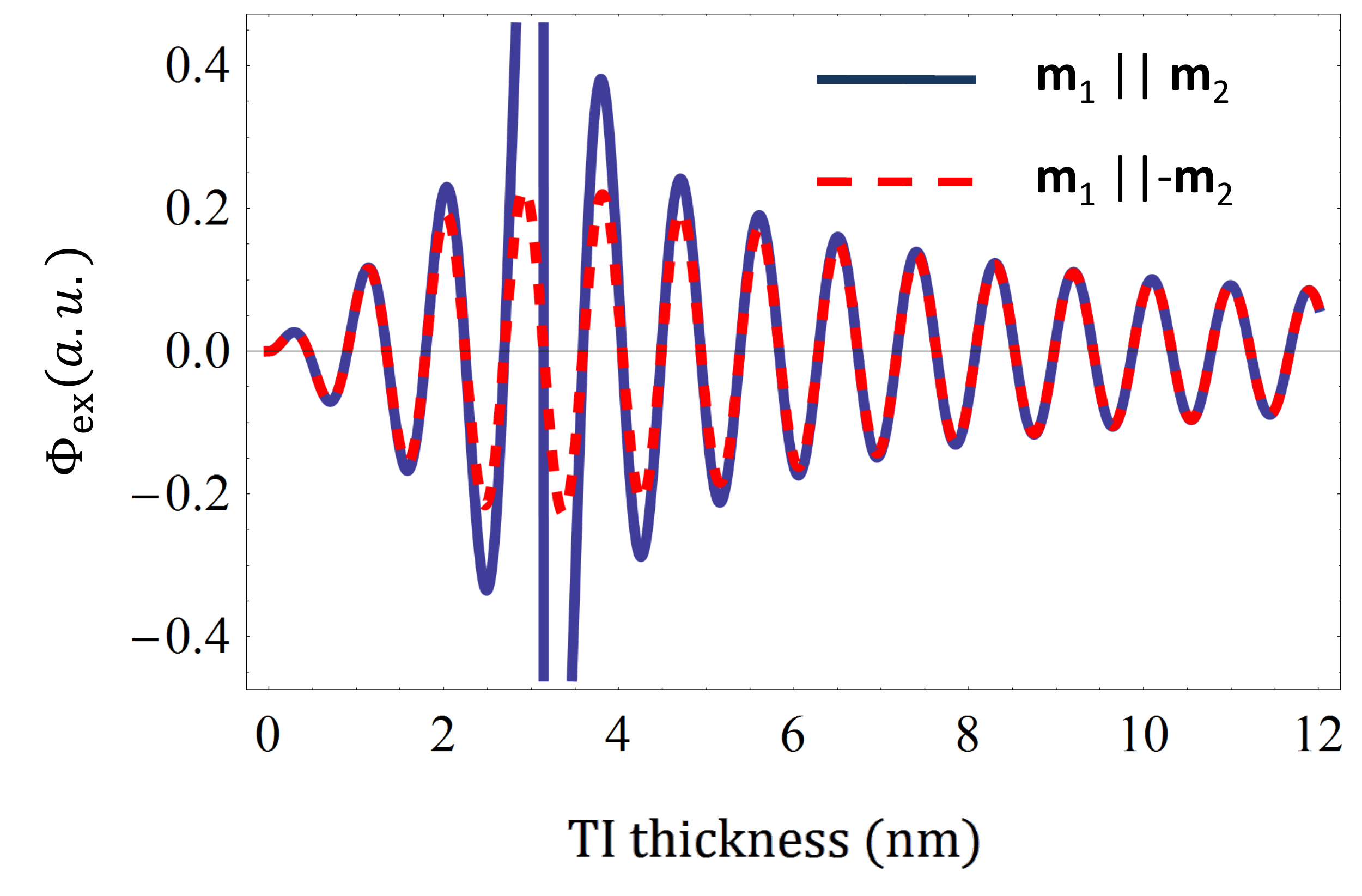}
 \caption{(Color online)  The interlayer exchange coupling ($\Phi_\mathrm{ex}$)  as a function of the TI thickness ($t_{TI})$ at $T=0$ and $E_F=0$.  The exchange coupling is oscillatory with a maximum amplitude at $t_{TI}\approx 3\ \mathrm{nm}$. Other parameters: $J=0.1$\ eV, $B_1= 0.1\ \mathrm{eV\ nm^2}$, $k_{F\bot}=5\ \mathrm{nm^{-1}}$. \label{Fig3}}
\end{figure}

First, we see that the exchange coupling has the oscillatory nature of RKKY-type coupling (see Fig. \ref{Fig3}). However, its amplitude does not monotonically decrease with increasing thickness, but reaches its peak at a thickness $t_{crit}=\pi\sqrt{B_1/J}$ such that $\mathrm{\Delta }=J$. This trend is in accordance with the behavior of the spin susceptibility discussed above. The compensation between paramagnetic Dirac surfaces and diamagnetic gapped-surfaces leads to the diminishing magnetic susceptibility, and so the diminishing interlayer exchange coupling. For typical parameters of $\mathrm{Bi_2Se_3}$ thin films, the thickness for maximum interlayer exchange coupling is estimated to be in the range of 3-5 nm. 

\section{Conclusion}
In this work, we have shown that topological insulator films as spacers in spin-valve structures play a fruitful role in mediating the exchange coupling between two ferromagnets. In one hand, the TI film provides electrons for mediating the coupling, which induce stronger coupling for thinner film as the magnetic moment transfer rate is stronger. On the other hand, thin TI film will encounter gap-opening in its surface states due to the hybridization, which in turn suppresses the spin susceptibility and RKKY coupling. Therefore, our work provides a guide to optimize TI-based spin valves for spintronics application.

\begin{acknowledgments}
We acknowledge the financial support of MOE Tier II grant MOE2013-T2-2-125 (NUS Grant No. R-263-000-B10-112), and the National Research Foundation of Singapore under the CRP Programs “Next Generation Spin Torque Memories: From Fundamental Physics to Applications” NRF-CRP12-2013-01 and “Non-Volatile Magnetic Logic and Memory Integrated Circuit Devices” NRF-CRP9-2011-01.
\end{acknowledgments}

%

\end{document}